\documentclass[sigconf]{acmart}

\usepackage{booktabs}
\usepackage{multirow}
\usepackage{algorithm}
\usepackage{algpseudocode}
\usepackage{circledsteps}
\usepackage{tabularx}
\usepackage{booktabs}

\AtBeginDocument{%
  }

\algnewcommand{\IfThenElse}[3]{
  \State \algorithmicif\ #1\ \algorithmicthen\ #2\ \algorithmicelse\ #3}

\newcommand{\F}{\ensuremath{{{\mathbb{F}}}}}
\newcommand{\eq}[1]{\tilde{\ensuremath{\mathrm{eq}}}#1}

\definecolor{rowgray}{gray}{0.95}

\copyrightyear{2025}
\acmYear{2025}
\setcopyright{rightsretained}
\acmConference[HASP 2025]{Hardware and Architectural Support for Security and Privacy 2025}{October 19, 2025}{Seoul, Republic of Korea}
\acmBooktitle{Hardware and Architectural Support for Security and Privacy 2025 (HASP 2025), October 19, 2025, Seoul, Republic of Korea}
\acmPrice{}
\acmDOI{10.1145/3768725.3768737}
\acmISBN{979-8-4007-2198-4/25/10}

\begin{document}

\title[MTU: The Multifunction Tree Unit for Accelerating Zero-Knowledge Proofs]{MTU: The Multifunction Tree Unit for Accelerating Zero-Knowledge Proofs}

\author{Jianqiao Mo}
\affiliation{
    \institution{Tandon School of Engineering \\ New York University}
    \city{Brooklyn}
    \state{NY}
    \country{USA}
}
\email{jm8782@nyu.edu}

\author{Alhad Daftardar}
\affiliation{
    \institution{Tandon School of Engineering \\ New York University}
    \city{Brooklyn}
    \state{NY}
    \country{USA}
}
\email{ajd9396@nyu.edu}

\author{Joey Ah-Kiow}
\affiliation{
    \institution{Tandon School of Engineering \\ New York University}
    \city{Brooklyn}
    \state{NY}
    \country{USA}
}
\email{ja4844@nyu.edu}

\author{Kaiyue Guo}
\affiliation{
    \institution{Tandon School of Engineering \\ New York University}
    \city{Brooklyn}
    \state{NY}
    \country{USA}
}
\email{kg3209@nyu.edu}

\author{Benedikt B{\"u}nz}
\affiliation{
    \institution{Courant Institute \\ New York University}
    \city{New York}
    \state{NY}
    \country{USA}
}
\email{bb@nyu.edu}

\author{Siddharth Garg}
\affiliation{
    \institution{Tandon School of Engineering \\ New York University}
    \city{Brooklyn}
    \state{NY}
    \country{USA}
}
\email{sg175@nyu.edu}

\author{Brandon Reagen}
\affiliation{
    \institution{Tandon School of Engineering \\ New York University}
    \city{Brooklyn}
    \state{NY}
    \country{USA}
}
\email{bjr5@nyu.edu}

\renewcommand{\shortauthors}{Mo et al.}

\begin{abstract}
Zero-Knowledge Proofs (ZKPs) are critical for privacy-preserving techniques and verifiable computation.
Many ZKP protocols rely on key kernels such as the SumCheck protocol and Merkle Tree commitments to enable their key security properties.
These kernels exhibit balanced binary tree computational patterns, which enable efficient hardware acceleration.
Although prior work has investigated accelerating these kernels as part of an overarching ZKP protocol, exploiting this common tree pattern remains relatively underexplored.
We conduct a systematic evaluation of these tree-based workloads under different traversal strategies, analyzing performance on multi-threaded CPUs and the Multifunction Tree Unit (MTU) hardware accelerator.
We introduce a hardware-friendly Hybrid Traversal for binary tree 
that improves parallelism and scalability while significantly reducing memory traffic on hardware.
Our results show that MTU achieves up to $1478\times$ speedup over CPU at DDR-level bandwidth and that our hybrid traversal outperforms breadth-first search by up to $3\times$. 
These findings offer practical guidance for designing efficient hardware accelerators for ZKP workloads with binary tree structures.

\end{abstract}



\maketitle

\section{Introduction}
\label{sec:intro}

Zero-Knowledge Proofs are cryptographic protocols that allow a prover to convince a verifier that a statement is true without revealing any additional information beyond the validity of the statement itself. 
ZKPs have emerged as a critical building block for privacy-preserving techniques and verifiable computation in both academia and industry. 
They have been widely adopted in privacy-focused blockchain platforms, such as Zcash~\cite{zerocash} and Mina~\cite{bonneau2020mina}, and are increasingly explored in applications such as privacy-preserving machine learning \cite{10.1145/3617232.3624852, mo2023towards}.  
As real-world deployments demand better performance and scalability, hardware-accelerated implementations have become increasingly important.

A key computational primitive in many modern ZKP protocols is the SumCheck protocol~\cite{thaler_proofs_args_zk}.
SumCheck plays a central role in a variety of ZKPs, including HyperPlonk~\cite{hyperplonk}, Spartan~\cite{spartan}, Hyrax \cite{cryptoeprint:2017/1132}, and Libra \cite{cryptoeprint:2019/317}. 
In order to support SumCheck 
in the HyperPlonk protocol, several subroutines are required: Build MLE, MLE Evaluation, Multiplication Tree, and Product MLE. 
These workloads are detailed in \autoref{sec:Binary Tree Workloads in ZKP}.
These subroutines share a common computational pattern: they can all be mapped onto balanced binary trees, where each internal node is a computation derived from its two children.
In addition to these SumCheck-related operations, Merkle Tree commitments are another binary-tree-based primitive frequently used in ZKPs and other cryptographic systems, where hash-based commitments are built in hierarchical tree structures.

Balanced binary trees can be computed using either breadth-first (BFS or level-ordered) or depth-first (DFS) 
strategies. 
While both are equivalent in algorithmic complexity, their behavior under hardware execution differs significantly due to resource constraints and memory hierarchy considerations. 
Prior research has primarily focused on theoretical aspects such as asymptotic complexity and software implementation~\cite{erickson2023algorithms}. 
However, designing hardware for these workloads requires careful consideration of several practical factors, including parallelism efficiency, memory traffic/bandwidth pressure, the complexity of control logic, the initiation interval in pipeline stages, and the ability to interface efficiently with upstream and downstream modules. 
These considerations are critical 
but 
relatively less studied in the binary tree workloads.

In this paper, we address this gap by systematically evaluating binary tree workloads under various traversal strategies.
We analyze the performance scalability of different traversal methods on 
CPU.
For hardware evaluation, we utilize the Multifunction Tree Unit (MTU) architecture~\cite{daftardar2025zkSpeed, daftardar2025zkphire} to assess the runtime and speedup of different traversal methods, 
and
explore the design space trade-offs between area and performance across a range of hardware configurations. 
{The MTU is integrated in zkSpeed~\cite{daftardar2025zkSpeed} to perform end-to-end proof generation.}
Additionally, we 
analyze
the hardware-friendly Hybrid traversal scheme that balances the benefits of both BFS and DFS, enabling efficient binary tree computation on MTU-based accelerators.
Our 
contributions are as follows:

\begin{itemize}
    \item We analyze the hardware-friendly Hybrid traversal strategy that enables efficient, simple-control parallelism while significantly reducing memory traffic overhead.
    \item We evaluate four representative ZKP workloads -- Build MLE, MLE Evaluation, Multiplication Tree, and Merkle Tree Commitment -- using different traversal methods.
    \item We perform a systematic hardware evaluation across memory bandwidth settings ranging from DDR to HBM, and complete a design space exploration of MTU. Our results show that MTU achieves a speedup of $1478\times$ over the CPU baseline with DDR-level bandwidth and that our Hybrid traversal outperforms BFS by nearly $3\times$.
    \item Our results offer insights into how traversal strategies affect runtime, bandwidth usage, scalability, and area efficiency, providing practical guidance for the hardware design of future ZKP accelerators.
\end{itemize}

These findings contribute to a deeper understanding of how to design and optimize hardware for cryptographic workloads with tree-based dataflows.
The MTU accelerator 
demonstrates a compact hardware footprint.
This efficiency enables MTU to be integrated into larger SoC designs or chiplet-based systems, sharing a common HBM PHY for memory access. 
It can be flexibly applied to future SumCheck accelerators or reused as a polynomial commitment engine across diverse ZKP stacks.

\section{Background}
\label{sec:Background}

\subsection{ZKP and zkSNARKs}

ZKPs are a privacy-preserving technology that allow a prover to convince 
a verifier that they know some information such that a statement is true. 
The state-of-the-art ZKPs are Zero-Knowledge Succinct Non-interactive ARgument of Knowledge (zkSNARK), which offer small proofs and fast verifier time at high prover costs. 

zkSNARKs are constructed from the combination of an interactive oracle proof (IOP) and a polynomial commitment scheme (PCS). 
The IOP provides a mechanism for the prover to encode a computation as an arithmetic circuit, and subsequently, as a set of polynomials whose evaluations can be efficiently checked through queries by the verifier.
SumCheck is a key kernel present in many recent IOPs due to its ability to scale efficiently with increasing problem complexity.
{Number-Theoretic Transform (NTT) is an alternative to SumCheck
used in some ZKP protocols.} 
The PCS, in turn, binds the prover to these polynomials and enables the verifier to check the correctness without requiring access to the full polynomials.
There are two common PCS kernels, Multi-Scalar Multiplication (MSM) and Merkle Tree.
Together, IOPs and PCSs form the foundation of zkSNARK protocols such as HyperPlonk~\cite{hyperplonk} and Orion\cite{orion}.
We target workloads related to SumCheck and Merkle Tree, seen in HyperPlonk and Orion respectively, {due to their shared binary tree computation patterns suited for similar hardware acceleration.}

\subsection{SumCheck and MLEs}

SumCheck \cite{thaler_proofs_args_zk} is an interactive  protocol that allows a prover to convince a verifier that the sum of a multivariate polynomial, $f(x_1, \ldots, x_\mu)$, over 
the
\textit{boolean hypercube} is some claimed sum, 
$S$:
$S = \sum_{x_1 \in \{0,1\}} \sum_{x_2 \in \{0,1\}} \cdots \sum_{x_\mu \in \{0,1\}} f(x_1, x_2, \dots, x_\mu)$.
The boolean hypercube is equivalent to a $\mu$-bit space $\vec{x} \in \{0,1\}^\mu$.
At a high level, for a $\mu$-variable polynomial $f$, there are $\mu$ rounds of the protocol; in round $i$, the prover makes a claim about the variable $x_i$. 
In protocols like HyperPlonk, a key constraint about $f$ is that it is \textit{multilinear}, meaning each variable's maximum degree in $f$ is one. 
A general representation of multilinear polynomials, for example, with $\mu=3$, 
\begin{equation}
\label{eq:multilinear_poly}
\begin{aligned}
    f(x_1, x_2, x_3) = f(0,0,0)(1-x_1)(1-x_2)(1-x_3) ~+ \\ 
    f(0,0,1)(1-x_1)(1-x_2)x_3+ \cdots \\
     + f(1,1,1) x_1 x_2 x_3
\end{aligned}
\end{equation}
This representation defines polynomials using evaluations at specific input points (e.g., $f(0,0,1)$) rather than using coefficients. 
Additionally, the points can be stored as a \textit{lookup table}, where each variable $x_i$ can be interpreted as a bit. 
For example, $f(0, 0, 0)$ can be stored at index 0 in the table; $f(0, 1, 0)$ is stored at index 2.

Throughout SumCheck rounds, $f$ 
will be evaluated at non-Boolean values of $x_i$,
even though $f$ is originally defined (and stored in hardware) over Boolean inputs.
This is done by extrapolating from values stored in the lookup table.
These values are called \textit{extensions}, thus
$f$ is called a \textit{multilinear extension} (MLE).
The term \textit{MLE table} denotes the corresponding lookup table structure mentioned above, where each entry's {data value (typically) occupies ~256} bits in ZKP applications.

In protocols like HyperPlonk, the polynomials upon which SumChecks are performed are often the result of tree-based compute patterns (e.g., Build MLE in  \autoref{sec:build_mle}).
Therefore, a fast tree unit is essential to ensure these computational stages do not become a performance bottleneck in SumCheck-based protocols.

\subsection{Merkle Tree Commitment}

Merkle trees in ZKP allow a prover to commit to a large vector of data using a compact root hash. 
Each leaf node of the tree corresponds to a hashed element in the original vector, while each internal node is the
cryptographic 
hash 
(e.g., SHA3~\cite{SHA3keccak}, Poseidon Hash~\cite{poseidon})
of its two child nodes. 
This recursive construction culminates in a single root hash that serves as a commitment to the entire vector.
Unlike MSM-based PCS constructions, Merkle trees do not require elliptic curve operations and thus typically offer faster prover runtime.
However, this comes at the cost of increased proof sizes and longer verifier latency.
Merkle-based PCS protocols, such as Orion~\cite{orion}, leverage this tradeoff by targeting environments where prover speed is critical and verifier costs are amortized or less constrained.
Merkle tree construction naturally maps to a binary tree reduction pattern,
which benefits from an efficient tree hardware unit, especially for vectors of lengths $\geq 2^{20}$.

\section{Binary Tree Traversal}
\label{sec:computation}

\subsection{Binary Tree Workloads in ZKP}
\label{sec:Binary Tree Workloads in ZKP}
Balanced binary tree computations are prevalent in ZKP protocols, particularly in SumCheck-based constructions~\cite{hyperplonk, spartan}.
These computation patterns appear repeatedly across ZKP applications and often form performance bottlenecks. 
Fortunately, their shared structure allows the design of a general hardware module that can accelerate them all with minimal specialization. 
zkSpeed introduces the MTU as a module capable to support multiple binary-tree-based computations. 
In this section, we highlight several common binary tree computations in ZKPs that motivate the design of the MTU.

\subsubsection{Build MLE}
\label{sec:build_mle}
Build MLE is an important subroutine in SumCheck related protocols.
SumCheck allows the prover to prove the sum of a 
multivariate 
polynomial $f$ 
over the Boolean HyperCube 
is zero, but in many ZKP protocols,
the goal could be stronger: to prove that $f(\vec{x}) = 0$ for every Boolean HyperCube input, i.e., each lookup table entry evaluates to 0.
However, running a SumCheck directly on $f(\vec{x})$ with a claimed sum $S = 0$ is not sufficient, because
two terms might be nonzero yet cancel each other out, meaning the total sum is zero even if individual evaluations are not zero.

To ensure correctness,
a 
polynomial $\eq(\vec{x}, \vec{r})$ is introduced \cite{blumberg2014verifiable, hyperplonk, spartan, setty2020quarks}:  
\:\:$\eq(\vec{x}, \vec{r}) = \prod_{i=1}^{\mu}\big(r_i x_i+(1-r_i)(1-x_i)\big)$,\:\:
where $\vec{x} \in \{0,1\}^\mu$, and $\vec{r} \in \F^\mu $ is a $\mu$-length vector of random values (challenges) in a finite field $\F$.
By proving $\sum_{\vec{x} \in \{0,1\}^\mu} \big(f(\vec{x}) \cdot \eq(\vec{x}, \vec{r})\big) =0$ with SumCheck, it implies that $f(\vec{x}) =0$ $\forall \vec{x} \in \{0,1\}^\mu$.
Its correctness is shown in \cite{spartan,campanelli2019legosnark} and is known as a ZeroCheck in~\cite{plonk, hyperplonk}. 
We denote the computation of $\eq(\vec{x}, \vec{r})$ as \textit{Build MLE}.

\begin{figure}[t!]
\centerline{\includegraphics[width=0.9\columnwidth]{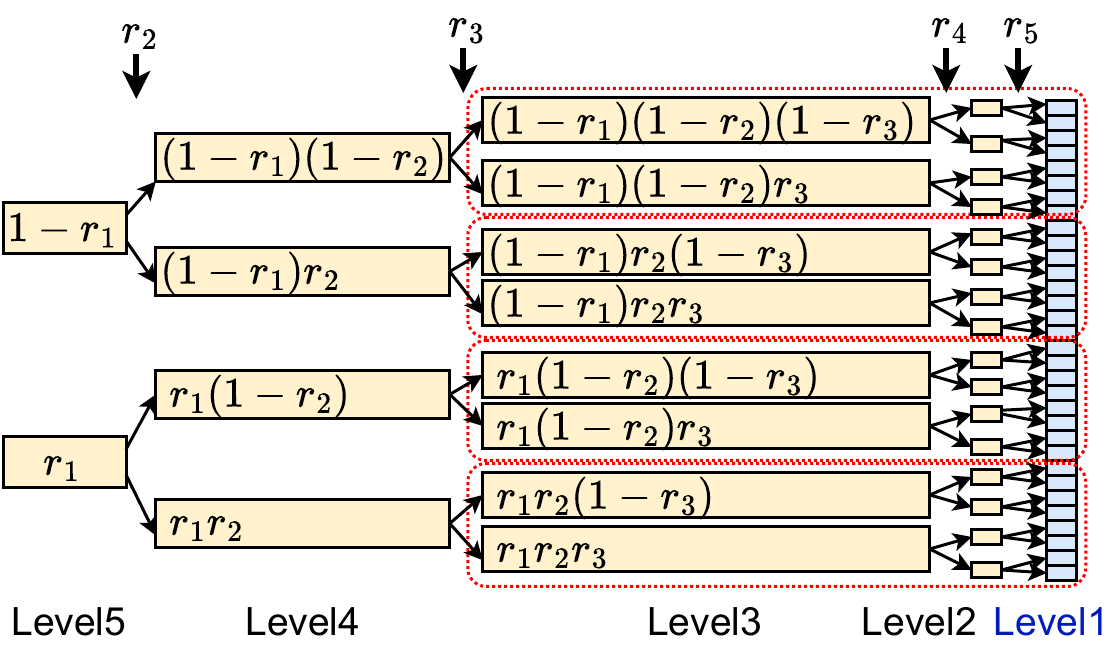}}
\caption{Dataflow for computing Build MLE (output size $2^5$).
The result level (blue) is indexed as Level 1.
Partitioning into four subtrees is highlighted with red dashed boxes to illustrate parallel DFS processing.
}
\label{fig:build_mle}
\end{figure}

To perform ZeroCheck, the prover must evaluate $\eq(\vec{x}, \vec{r})$ at all $\vec{x}\in \{0,1\}^\mu$, which in fact is a list of products from $(1-r_1)(1-r_2)\cdots(1-r_{\mu})$ to $r_1r_2\cdots r_{\mu}$.
A naive approach requires $(\mu-1) 2^{\mu}$ modular multiplications to compute all such products individually.
However, this cost can be reduced substantially by computing the values via a binary tree structure, as illustrated in \autoref{fig:build_mle}. 
The binary tree method reduces the number of modular multiplications to $2^{\mu+1} - 4$, providing significant computational savings. 
This optimization motivates the tree-based hardware design capable of efficiently supporting this pattern.
In the tree-based computation, we can reduce one modular multiplication by using:
$(1-r_1)\cdot (1-r_2)=(1-r_1)-(1-r_1)\cdot r_2$.
That is, $(1-r_1)(1-r_2)$ can be computed from $(1-r_1)r_2$, instead of computing an extra modular multiplication with $(1-r_2)$ and $(1-r_1)$.

\subsubsection{MLE Evaluation}
MLE Evaluation is also an important step in SumCheck-related ZKP protocols. 
It is used to evaluate a {multilinear} polynomial on a given challenge point.

Generally, there is a multilinear polynomial $f(\vec{x})$, and it is depicted by its values on the Boolean Hypercube (i.e., $f(\vec{x})$ $\forall \vec{x} \in \{0,1\}^\mu$), in the form of a look-up table (see \autoref{sec:Background}). 
The prover will be asked to evaluate $f$ on a random challenge $\vec{r} \in \F^{\mu}$, i.e., to compute $f(r_1, r_2, \cdots, r_{\mu})$.
Since $f$ is a multilinear polynomial,
as shown in \autoref{eq:multilinear_poly},
the MLE Evaluation can be written as

\begin{equation}
\label{eq:MLE_eval}
\begin{aligned}
    f(r_1, \cdots, r_\mu) = \big(f(0,\cdots,0)(1-r_1)\cdots(1-r_\mu)\big) + 
     \cdots \\
     +~\big(f(1,\cdots,1) x_1 \cdots x_\mu\big)
\end{aligned}
\end{equation}

A naive method is to modular-multiply every individual term and sum all at the end.
Similar to Build MLE, we can substantially improve the computation by following the binary tree structure, while the tree is now inverted (reduction computation, as shown in \autoref{fig:MLE_eval}).
With the binary tree structure, the number of modular multiplications can be optimized from $\mu \cdot 2^{\mu}$ to $2^{\mu}-1$. 
In \autoref{fig:MLE_eval} we applied this trick to further reduce one modular multiplication:
$f_0\cdot (1-r_1)+ f_1 \cdot r_1 = f_0 + r_1 \cdot (f_1-f_0)$.
Since modular addition is cheaper than modular multiplication, we can replace the original form on the left of the equation with one mod-mul on the right.

\subsubsection{Merkle Tree}
\label{sec:merkle tree workload}
The Merkle tree is a widely used cryptographic commitment scheme in which the root of a binary tree serves as the commitment to a set of leaf values. 
Individual leaves can later be revealed and efficiently verified as part of the committed structure using authentication paths. 
In ZKP protocols, Merkle trees are often employed as hash-based polynomial commitment schemes.

\begin{figure}[t!]
\centerline{\includegraphics[width=1\columnwidth]{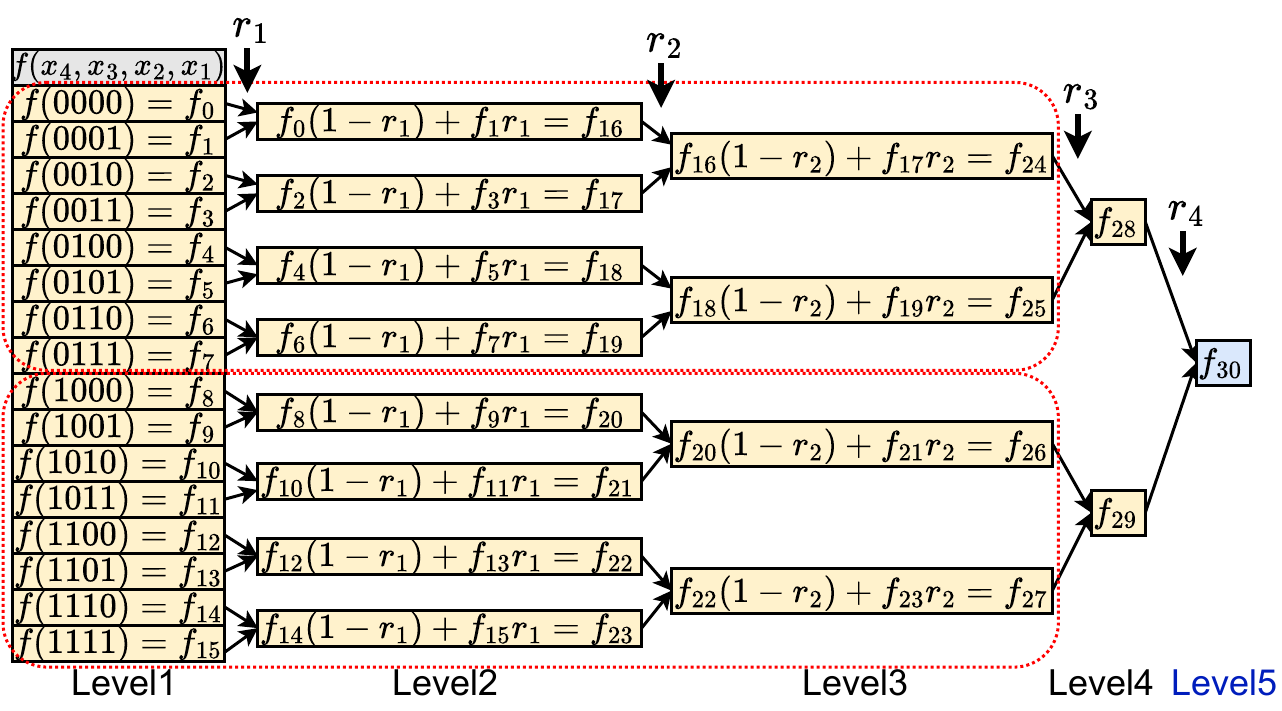}}
\caption{Dataflow for computing the MLE evaluation workload (input size $2^4$). 
Final output is marked in blue.
Partitioning into two subtrees is shown with red dashed boxes to illustrate parallel DFS processing.
}
\label{fig:MLE_eval}
\end{figure}

Unlike MLE Evaluation in \autoref{eq:MLE_eval}, the Merkle Tree naturally has a binary tree structure.  
The Merkle tree commitment workload follows a similar computation pattern as illustrated in \autoref{fig:MLE_eval}. 
The primary difference lies in the operation at each node: Level 1 consists of the hash values of the original input data, and each subsequent level is computed by hashing pairs of nodes from the previous level. 
Instead of modular arithmetic, each tree node performs a cryptographic hash 
of its two child nodes, forming a hierarchical hash structure that culminates in a single root commitment.
This structural similarity enables reuse of binary-tree acceleration strategies for Merkle tree construction and verification within zero-knowledge systems.

\subsubsection{Multiplication Tree and Product MLE}
\label{sec:prod_mle}
Binary tree-style product computations are not limited to SumCheck protocols; many cryptographic algorithms require computing products of the form 
$\prod_{i=0}^{2^\mu-1} f_i$. 
In cryptographic contexts, these products typically involve modular multiplication, which is significantly more expensive than standard integer multiplication. 
A simple way is to accumulate the product one by one, which will incur high latency costs because of the data dependency. 
Therefore, computations using binary tree structures must be optimized to reduce computational cost.

One notable example is the \textit{Product MLE} computation in HyperPlonk, which resembles a standard multiplication tree. 
However, unlike a typical tree that only outputs the final product at the root, the Product MLE computation requires the outputs from \emph{all} intermediate levels of the tree. 
These intermediate values are essential for downstream steps in the protocol. 
The multiplication tree and Product MLE workloads follow a similar structural pattern to that shown in \autoref{fig:MLE_eval}, with the primary difference in the node operation. 
Instead of modular addition and multiplication, each simply computes a modular multiplication of two nodes from the previous level. 
Product MLE's requirement of outputting the intermediate results (levels starting from Level 2) increases bandwidth demands.

\textbf{SumCheck}. 
At each round $i$ of the SumCheck protocol, one can observe that the summation $\sum_{x=0}^{2^i - 1} f(x)$ follows a tree-based reduction pattern. 
However, unlike modular multiplication, modular addition is relatively inexpensive and incurs minimal latency (e.g., in zkSpeed \cite{daftardar2025zkSpeed}, mod-add is a one stage pipeline).
As a result, the accumulation dependency does not significantly impact the throughput or total latency. 
An accumulator is therefore sufficient to compute the sum efficiently, without the need for a 
binary reduction tree.
{For product computation, accumulator method would significantly influence the latency because of long mod-mul latency.}

\subsection{General Binary Tree Traversals}

Binary tree computations can be traversed in various ways, with the two most common strategies being breadth-first search (BFS, or level-order traversal) and depth-first search (DFS, including inorder, preorder, postorder traversal). 
In DFS, computation progresses by recursively visiting one branch of the tree down to its leaves before backtracking. 
In contrast, BFS processes the tree level by level, evaluating all nodes at a given depth before moving to the next.

\begin{figure}[t]
\centerline{\includegraphics[width=1\columnwidth]{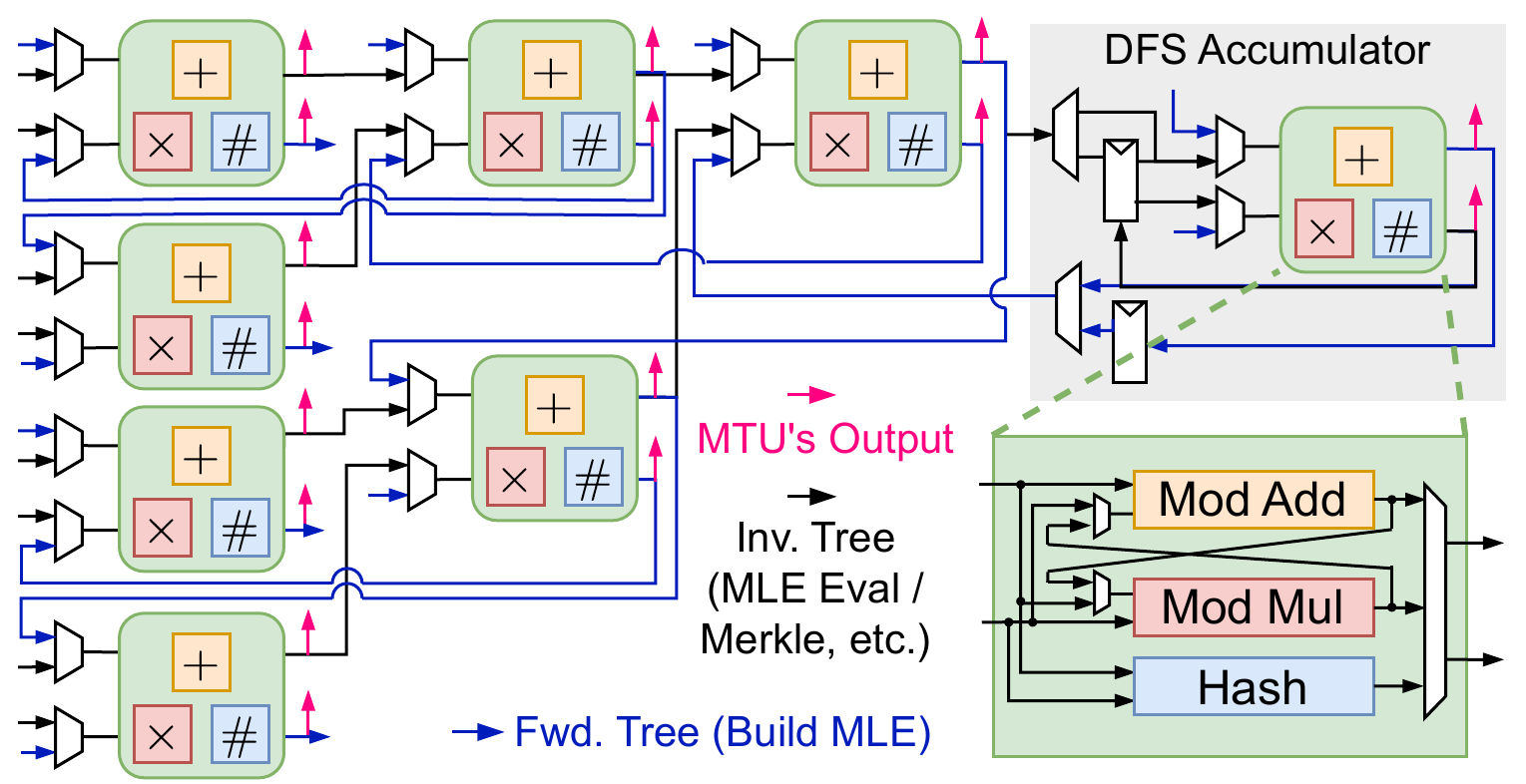}}
\caption{MTU architecture with 8 PEs for Hybrid Traversal.
Each PE supports modular arithmetic and hashing. Connection enables both forward and inverted binary trees computation. 
All PEs can directly output results of the accelerator.
}
\label{fig:MTU arch}
\end{figure}

\textbf{BFS (level-ordered)}.
Each traversal strategy offers trade-offs. 
BFS is straightforward to implement and control, making it a natural choice in software for binary tree workloads 
mentioned above.
Its dataflow typically involves reading all nodes at a given level from memory, performing computations in parallel, and then writing results back to memory. 
In the example of \autoref{fig:MLE_eval}, BFS computation follows Level 2, 3 and so on. 
This regular, level-by-level structure is amenable to parallelism.
However, BFS has significant hardware design challenges. 
If implemented entirely on-chip, it requires storing all intermediate levels of the tree, which can incur prohibitive memory costs. 
For example, processing Build MLE of $2^{23}$ workload size 
would require 128 MB of on-chip SRAM, and it grows linearly with the size.
This is inefficient for typical ASIC designs. 
Alternatively, if intermediate levels are stored off-chip, the design becomes bandwidth-limited, and the achievable parallelism is constrained by memory access throughput.
These limitations motivate the need for more memory-efficient traversal strategies that balance parallelism with resource constraints.
Hardware accelerators like Nocap~\cite{nocap} applied BFS for their Merkle tree computations.

\textbf{DFS}.
DFS offers substantial memory and bandwidth savings compared to BFS through on-chip data reuse.
Since DFS computes each branch to completion before moving on, intermediate results are consumed immediately and do not need to be stored for long durations. 
For example in \autoref{fig:MLE_eval}, it follows the order $f_0$, $f_1$ to $f_{16}$, and then $f_2$, $f_3$, $f_{17}$, $f_{24}$, and so on.
Due to the data dependency, DFS is naturally less compatible with parallel hardware execution. 
A naive approach (i.e., feeding instructions in DFS order to multiple PEs)
results in complex scheduling and high latency from 
deep recursion.
{Since PEs prioritize deeper levels, inputs are stalled, causing irregular arrival rates and an initiation interval (II) greater than one, where not all PEs can accept new inputs every cycle.}

Another way to simplify 
parallelized DFS
is to divide the tree into multiple subtrees and assign each subtree to a dedicated PE. 
In \autoref{fig:MLE_eval}, we divide the workload into two subtrees,
which enables two parallel PEs to compute each subtree, and their results will be merged at the end.
\autoref{fig:build_mle} is similar, while Level 4 will be preprocessed.
This improves concurrency, but since the tree is divided into several subtrees, the initial input or output indices are no longer sequential.
If an upstream module (e.g., another accelerator block) is producing tree inputs in \textit{strict} index order, such DFS traversal cannot be directly pipelined with it.
This can be found in zkSpeed \cite{daftardar2025zkSpeed} where the Fraction MLE feeds the Product MLE unit in the index order.
These challenges highlight the need for traversal strategies that preserve DFS's reuse efficiency while improving input regularity and pipelining compatibility.
Hardware accelerators like UniZK~\cite{unizk} apply a similar philosophy of dividing trees, but within each subtree it still performs BFS individually.

\begin{table}[t]
\centering
\caption{Different binary tree traversal methods.
}
\resizebox{0.98\columnwidth}{!}{
\setlength{\tabcolsep}{2mm}{
\begin{tabular}{l|lll}
\hline
\textbf{Traversal}           & \textbf{BFS}    & \textbf{DFS}    & \textbf{MTU Hybrid} \\ \hline
\textbf{Time Complexity}     & $O(n)$          & $O(n)$          & $O(n)$                \\
\textbf{Parallel Control}    & Easy            & Hard            & Medium                \\
\textbf{Initiation Interval} & =1              & \textgreater{}1 & =1                    \\
\textbf{Input indices}       & Continuous      & Discont.   & Continuous            \\
\textbf{Memory Cost}         & High            & Low             & Low                   \\
\textbf{Bandwidth Cost}           & Linear to \#PEs & Fixed           & Fixed                 \\ \hline
\end{tabular}
}
}
\label{tab:compare_traversal}
\end{table}

These traversal strategies 
are summarized in \autoref{tab:compare_traversal}. 
Regardless of the traversal method, the asymptotic time complexity remains the same $O(n)$ since each approach must visit all $n$ nodes in the tree once \cite{erickson2023algorithms}.
Ideally, parallelism may improve this latency to $O(n/p)$ for $p$ parallel PEs. 
The key differences lie in space complexity and their suitability for parallel execution. 
BFS enables straightforward parallelism across tree levels but incurs a space complexity of $O(n)$ due to the need to store a level of intermediate data. 
In contrast, DFS has lower memory requirements because intermediate results can be reused and discarded quickly, but it presents greater challenges in coordinating parallel execution and achieving efficient pipelining.

\section{Hardware-Friendly Hybrid Traversal} 
\label{sec:Hybrid Traversal}

To balance the trade-offs between space complexity and parallelism, MTU introduces a hybrid traversal strategy tailored for hardware efficiency. 
This method combines the low-memory advantages of DFS with the parallelism-friendly nature of BFS.

The starting point 
is rate-matching the upstream. 
For an inverted binary tree in \autoref{fig:MLE_eval}, if eight nodes of Level 1 arrive per cycle, four PEs will be allocated to process them in parallel, and two PEs are used for the second level, and one PE for the third level. 
As shown in \autoref{fig:MTU arch}, these seven PEs are connected to form a pipeline, enabling the first to third levels to be grouped, 
and processed by the seven PEs pipeline.
The pipeline generates one output per cycle at the fourth level, which then can be processed by a single PE 
in DFS method
(\autoref{sec:DFS_Accumulator}).
Thus, the traversal is divided: 
Level 1-3 are grouped, handled in level-order to match input rate and ensure continuous indexing.
Intermediate data are reused within the pipeline.
On the other hand, the deeper levels 
($\geq$ Level 4)
adopt a depth-first strategy which has low space complexity.
{This hybrid scheme no longer stores the entire intermediate levels.
It 
balances parallelism and on-chip reuse
with simple scheduling.
Furthermore, the design is scalable,
as the number of PEs can be adjusted to match an input rate,
forming a full pipeline with other 
modules.
}

\subsection{MTU Architecture}
\label{sec:MTU_Architecture}

\autoref{fig:MTU arch} illustrates the architecture of MTU.
The MTU is composed of several PEs that are connected to support both standard (forward) and inverted binary trees. 
The inverted tree mode 
(workloads like Merkle tree commit) 
MTU can accept eight inputs per cycle. 
That is, the PEs will take eight elements of Level 1 (e.g., $f_{0}$ to $f_{7}$ in \autoref{fig:MLE_eval}) into the pipeline each cycle, and produce a Level 4 element (e.g., $f_{28}$) into the last PE.
The architecture can be extended on demand.
Putting more PEs in the front allows the MTU to process more inputs per cycle. 
Each PE is equipped with a SHA3 engine to accelerate hash-based operations required in Merkle Tree, along with modular addition and multiplication units to support arithmetic workloads 
(e.g., MLE Evaluation and Multiplication Tree).

\subsubsection{DFS Accumulator}
\label{sec:DFS_Accumulator}
We show the scheduling of MTU's DFS Accumulator in \autoref{tab:DFS_scheduling_inv_tree},
assuming an MTU configuration as \autoref{fig:MTU arch} and the PE latency is one cycle. 
As mentioned above, 
since Level 4 elements come at one per cycle, Level 5 outputs elements every two cycles, Level 6 outputs elements every four cycles, and so on.
The temporary results (e.g., $L5_0$) are buffered in a small SRAM and consumed once their corresponding input pairs are ready.
This allows interleaving across levels during scheduling. This scheduling is determined by the generation rate of each level. A simple controller can handle this efficiently by prioritizing deeper levels: whenever the inputs of a deeper level are available, that level is triggered to compute.
At the beginning, there are unavoidable gaps in the schedule, as each input pair for Level 5 and beyond requires multiple cycles to become ready. 
However, once multiple levels are processed, these gaps are filled. 
Each PE is also connected to the output ports (red wires in \autoref{fig:MTU arch}) to support Product MLE computation, which requires all elements of every level.

In the forward binary tree (Build MLE workload), the MTU can generate up to eight outputs per cycle. 
PEs are connected in blue wires in \autoref{fig:MTU arch}.
Unlike the inverted binary tree, here the four leftmost PEs generate the output level (Level 1).
Nodes from Level 3 to 1 are grouped 
and processed by the seven PEs pipeline.
Following the similar DFS scheme, the 
accumulator prepares Level 4 and above,
with its schedule shown in
\autoref{tab:DFS_scheduling_fwd_tree}.
Level 4 nodes are produced at an average rate of one per cycle, 
which allows Level 3 computations to launch every cycle and enables a continuous stream of eight Level 1 outputs.
{The rate of Level 4 determines the scheduling of higher levels.
Although there are gaps in the initial cycles, they will be filled as computation proceeds.
These gaps are dictated by the launch rates: for example, Level 5 launches every two cycles, while Levels 7 and 8 launch every eight and sixteen cycles, respectively. 
The scheduler inserts gaps at cycles 2 and 3 to ensure that, after sixteen cycles, the next Level 8 ($L8_1$) 
will not be blocked by overlap.}

\begin{table*}[t]
\centering
\caption{Scheduling of MTU DFS accumulator when processing an inverted binary tree.
$L4_{0}$ means the index-0 node of Level 4.
The PE 
takes two inputs from previous level, and produces one output at the next level.
}
\resizebox{2.05\columnwidth}{!}{
\setlength{\tabcolsep}{1mm}{
\begin{tabular}{lccccccccccccccccccccccccccccc}
\hline
\textbf{Cycle}   & 0 & 1        & 2                               & 3        & 4                               & 5        & 6                               & 7                               & 8                               & 9        & 10                              & 11                              & 12                              & 13                              & 14                              & 15                              & 16                              & 17        & 18                              & 19                              & 20                              & 21                              & 22                               & 23                              & 24                               & 25                              & 26                               & 27                              & ... \\ \hline
\textbf{Input A} & - & $L4_{0}$ & -                               & $L4_{2}$ & -                               & $L4_{4}$ & {\color[HTML]{E69138} $L5_{0}$} & $L4_{6}$                        & -                               & $L4_{8}$ & {\color[HTML]{E69138} $L5_{2}$} & $L4_{10}$                       & {\color[HTML]{9900FF} $L6_{0}$} & $L4_{12}$                       & {\color[HTML]{E69138} $L5_{4}$} & $L4_{14}$                       & -                               & $L4_{16}$ & {\color[HTML]{E69138} $L5_{6}$} & $L4_{18}$                       & {\color[HTML]{9900FF} $L6_{2}$} & $L4_{20}$                       & {\color[HTML]{E69138} $L5_{8}$}  & $L4_{22}$                       & {\color[HTML]{0000FF} $L7_{0}$}  & $L4_{24}$                       & {\color[HTML]{E69138} $L5_{10}$} & $L4_{26}$                       & ... \\
\textbf{Input B} & - & $L4_{1}$ & -                               & $L4_{3}$ & -                               & $L4_{5}$ & {\color[HTML]{E69138} $L5_{1}$} & $L4_{7}$                        & -                               & $L4_{9}$ & {\color[HTML]{E69138} $L5_{3}$} & $L4_{11}$                       & {\color[HTML]{9900FF} $L6_{1}$} & $L4_{13}$                       & {\color[HTML]{E69138} $L5_{5}$} & $L4_{15}$                       & -                               & $L4_{17}$ & {\color[HTML]{E69138} $L5_{7}$} & $L4_{19}$                       & {\color[HTML]{9900FF} $L6_{3}$} & $L4_{21}$                       & {\color[HTML]{E69138} $L5_{9}$}  & $L4_{23}$                       & {\color[HTML]{0000FF} $L7_{1}$}  & $L4_{25}$                       & {\color[HTML]{E69138} $L5_{11}$} & $L4_{27}$                       & ... \\
\textbf{Output}  & - & -        & {\color[HTML]{E69138} $L5_{0}$} & -        & {\color[HTML]{E69138} $L5_{1}$} & -        & {\color[HTML]{E69138} $L5_{2}$} & {\color[HTML]{9900FF} $L6_{0}$} & {\color[HTML]{E69138} $L5_{3}$} & -        & {\color[HTML]{E69138} $L5_{4}$} & {\color[HTML]{9900FF} $L6_{1}$} & {\color[HTML]{E69138} $L5_{5}$} & {\color[HTML]{0000FF} $L7_{0}$} & {\color[HTML]{E69138} $L5_{6}$} & {\color[HTML]{9900FF} $L6_{2}$} & {\color[HTML]{E69138} $L5_{7}$} & -         & {\color[HTML]{E69138} $L5_{8}$} & {\color[HTML]{9900FF} $L6_{3}$} & {\color[HTML]{E69138} $L5_{9}$} & {\color[HTML]{0000FF} $L7_{1}$} & {\color[HTML]{E69138} $L5_{10}$} & {\color[HTML]{9900FF} $L6_{4}$} & {\color[HTML]{E69138} $L5_{11}$} & {\color[HTML]{6AA84F} $L8_{0}$} & {\color[HTML]{E69138} $L5_{12}$} & {\color[HTML]{9900FF} $L6_{5}$} & ... \\ \hline
\end{tabular}
}
}
\label{tab:DFS_scheduling_inv_tree}
\end{table*}

\begin{table*}[t]
\centering
\caption{Scheduling of MTU DFS accumulator when processing a forward binary tree (Build MLE).
$L7_{0}$ means the index-0 node of Level 7.
The PE 
takes an input from previous level, and produces two outputs at the next level.
}
\resizebox{2.05\columnwidth}{!}{
\setlength{\tabcolsep}{1mm}{
\begin{tabular}{lccccccccccccccccccccccccccccc}
\hline
\textbf{Cycle}    & 0                               & 1                               & 2 & 3 & 4                               & 5                               & 6                               & 7                               & 8 & 9                               & 10                              & 11                              & 12                              & 13                              & 14                              & 15                              & 16                              & 17                              & 18                              & 19                              & 20                              & 21                              & 22                              & 23                              & 24        & 25                              & 26                              & 27                               & ... \\ \hline
\textbf{Input}    & {\color[HTML]{6AA84F} $L8_{0}$} & -                               & - & - & {\color[HTML]{0000FF} $L7_{0}$} & -                               & {\color[HTML]{9900FF} $L6_{0}$} & -                               & - & {\color[HTML]{E69138} $L5_{0}$} & {\color[HTML]{9900FF} $L6_{1}$} & {\color[HTML]{E69138} $L5_{1}$} & {\color[HTML]{0000FF} $L7_{1}$} & {\color[HTML]{E69138} $L5_{2}$} & {\color[HTML]{9900FF} $L6_{2}$} & {\color[HTML]{E69138} $L5_{3}$} & {\color[HTML]{6AA84F} $L8_{1}$} & {\color[HTML]{E69138} $L5_{4}$} & {\color[HTML]{9900FF} $L6_{3}$} & {\color[HTML]{E69138} $L5_{5}$} & {\color[HTML]{0000FF} $L7_{2}$} & {\color[HTML]{E69138} $L5_{6}$} & {\color[HTML]{9900FF} $L6_{4}$} & {\color[HTML]{E69138} $L5_{7}$} & -         & {\color[HTML]{E69138} $L5_{8}$} & {\color[HTML]{9900FF} $L6_{5}$} & {\color[HTML]{E69138} $L5_{9}$}  & ... \\
\textbf{Output A} & -                               & {\color[HTML]{0000FF} $L7_{0}$} & - & - & -                               & {\color[HTML]{9900FF} $L6_{0}$} & -                               & {\color[HTML]{E69138} $L5_{0}$} & - & -                               & $L4_{0}$                        & {\color[HTML]{E69138} $L5_{2}$} & $L4_{2}$                        & {\color[HTML]{9900FF} $L6_{2}$} & $L4_{4}$                        & {\color[HTML]{E69138} $L5_{4}$} & $L4_{6}$                        & {\color[HTML]{0000FF} $L7_{2}$} & $L4_{8}$                        & {\color[HTML]{E69138} $L5_{6}$} & $L4_{10}$                       & {\color[HTML]{9900FF} $L6_{4}$} & $L4_{12}$                       & {\color[HTML]{E69138} $L5_{8}$} & $L4_{14}$ & -                               & $L4_{16}$                       & {\color[HTML]{E69138} $L5_{10}$} & ... \\
\textbf{Output B} & -                               & {\color[HTML]{0000FF} $L7_{1}$} & - & - & -                               & {\color[HTML]{9900FF} $L6_{1}$} & -                               & {\color[HTML]{E69138} $L5_{1}$} & - & -                               & $L4_{1}$                        & {\color[HTML]{E69138} $L5_{3}$} & $L4_{3}$                        & {\color[HTML]{9900FF} $L6_{3}$} & $L4_{5}$                        & {\color[HTML]{E69138} $L5_{5}$} & $L4_{7}$                        & {\color[HTML]{0000FF} $L7_{3}$} & $L4_{9}$                        & {\color[HTML]{E69138} $L5_{7}$} & $L4_{11}$                       & {\color[HTML]{9900FF} $L6_{5}$} & $L4_{13}$                       & {\color[HTML]{E69138} $L5_{9}$} & $L4_{15}$ & -                               & $L4_{17}$                       & {\color[HTML]{E69138} $L5_{11}$} & ... \\ \hline
\end{tabular}
}
}
\label{tab:DFS_scheduling_fwd_tree}
\end{table*}

{The MTU Hybrid Traversal, paired with the DFS Accumulator PE, enables deeper recursion with low memory overhead and 
fixed-pattern control.}
The design insight is to match the generation rate between levels: 
grouping Levels 1–3 across seven PEs, while the DFS Accumulator synchronizes with this group to maintain throughput.

\section{Methodology}
\label{sec:Methodology}

\textbf{Simulation and Workloads.}
We evaluate MTU using the representative workloads described in~\autoref{sec:Binary Tree Workloads in ZKP}. 
Software performance is measured on Intel Xeon Gold 5218 CPU~\cite{Intel_CPU} with Thermal Design Power (TDP) 125\,W, and DDR4 memory. 
The baseline software implementation is built on top of the HyperPlonk and the \texttt{arkworks} Rust library~\cite{hyperplonkespresso, arkworks}.
Software multi-threaded tests are handled by Rayon.
To explore scalability, we evaluate the CPU baseline with thread counts ranging from 1 to 32, and compare it against MTU configurations using 2 to 32 PEs.
Since MTU uses hybrid binary tree traversal with a DFS Accumulator, we start from a configuration with 2 PEs.
We do not evaluate Hybrid traversal on the CPU, as it lacks the tight pipeline structure of hardware PEs, making it infeasible for the precise scheduling required for the DFS Accumulator stage in Hybrid traversal.
To reflect realistic system memory conditions, we simulate MTU performance under various bandwidth constraints, ranging from 64\,GB/s (comparable to CPU systems with DDR4~\cite{DDR4, DDR4_crucial}, DDR5~\cite{micron_ddr5}) to 1024\,GB/s (HBM-class memory~\cite{hbm2, hbm3, haac, bts,moaccelerating}). 
This allows us to evaluate MTU's performance potential across both commodity and high-end memories.

\begin{figure*}[t]
\centerline{\includegraphics[width=2.15\columnwidth]{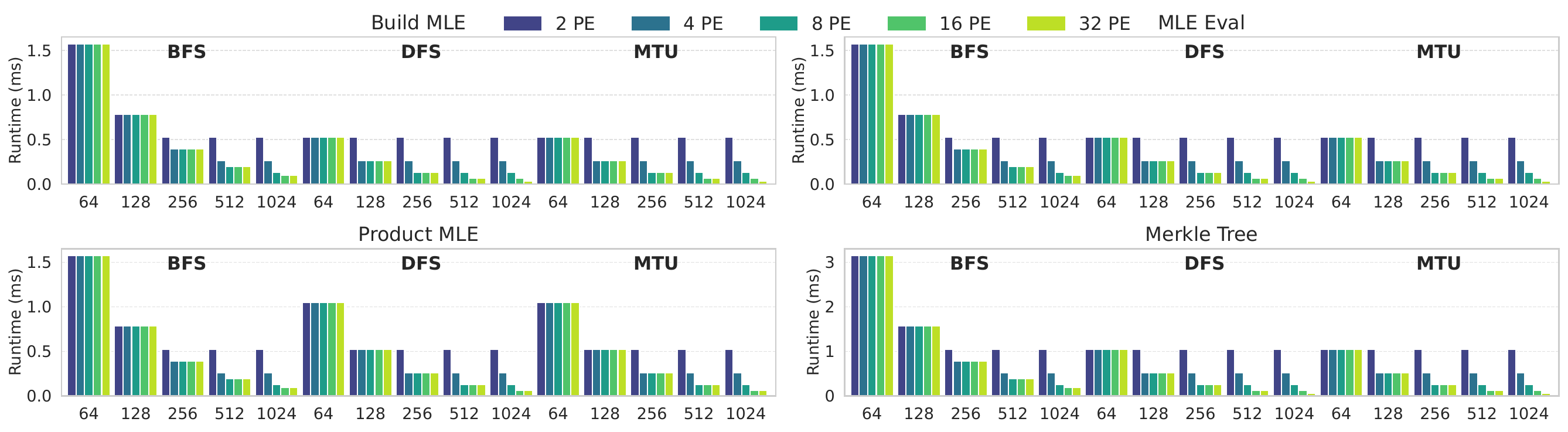}}
\caption{MTU runtime across workloads of size $2^{20}$, grouped by traversal types. 
Each group shows hardware runtime under varying hardware bandwidths (GB/s) and PE counts. 
Note that DFS is parallelized by partitioning the binary tree into subtrees, resulting in non-contiguous input/output indices.}
\label{fig:hardware_runtime}
\end{figure*}

\textbf{Hardware Implementation.}
We design a cycle-accurate simulator to model MTU's hybrid traversal, as discussed in~\autoref{sec:Hybrid Traversal}, together with a scheduler to control the task for the DFS accumulator in MTU. 
Our simulation uses 255-bit datatypes, consistent with the finite field setting in HyperPlonk.
We use Catapult HLS 2024 to generate fully-pipelined Montgomery multipliers supporting arbitrary prime moduli (following prior work~\cite{szkp, daftardar2025zkSpeed}). 
For hash computations in Merkle Tree, we instantiate the SHA3 block from OpenCores~\cite{opencores_sha3}. 
On-chip memory is modeled using SRAM 
generated by Synopsys 22\,nm Memory Compiler. 
Area and power figures are scaled to 7\,nm using technology scaling factors: $3.6\times$ for area, $3.3\times$ for power, and $1.7\times$ for delay~\cite{szkp, 22nm_TSMC_web, 7nm_TSMC_paper, 16nm_TSMC_paper,geng2023privacy}. 
The MTU is assumed to operate at a clock frequency of 1\,GHz.

\begin{figure}[t]
\centerline{\includegraphics[width=1.05\columnwidth]{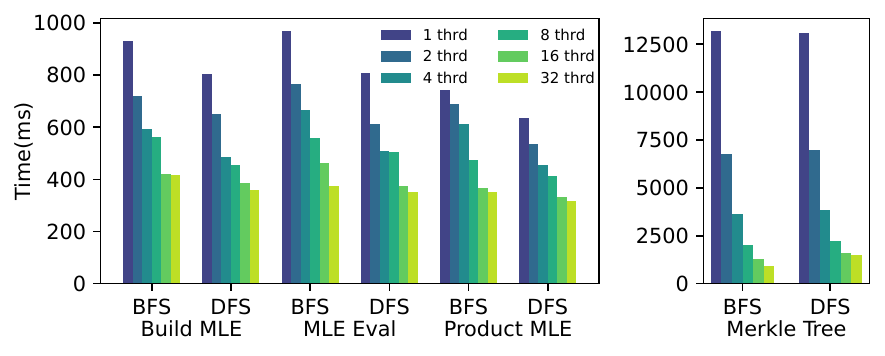}}
\caption{CPU performance across different workloads using BFS and DFS. Each workload is evaluated at size $2^{20}$.
}
\label{fig:CPU_runtime}
\end{figure}

\section{Evaluation}
\label{sec:evaluation}

Here, we evaluate each workload with different traversal methods on CPU and MTU.
We show performance scaling on CPU threads and hardware performance scaling with constrained bandwidth.

\subsection{Traversal Analyses on CPU}
\label{sec: CPU perf}

We first evaluate various binary tree traversal schemes on a CPU. 
\autoref{fig:CPU_runtime} presents the runtime performance for four representative workloads: constructing the Build MLE (resulting in an MLE table of size $2^{20}$ entries), evaluating the MLE, computing the Product MLE, and performing Merkle Tree commitment, each over an input table of size $2^{20}$. 
Under varying number of CPU threads, we compare traversal schemes and analyze how they benefit from parallelism.

Among the workloads, the Merkle Tree shows clear performance scaling with increased thread count. 
In contrast, workloads involving modular arithmetic -- such as Build MLE, MLE evaluation, and Product MLE -- do not scale linearly with threads. 
This difference stems from the nature of the underlying computation: the SHA3 hash used in the Merkle Tree is well-optimized on modern CPUs via intrinsic SHA Extensions~\cite{intel_sha3} and benefits significantly from thread-level parallelism, whereas modular arithmetic operations are computationally intensive with more computation dependency.

The performance difference between BFS and DFS traversal is not pronounced, especially for large, compute-heavy workloads like Merkle Tree commitment, where most of the execution time is dominated by computation.
The overhead introduced by traversal patterns becomes relatively minor in the overall execution time.
Nevertheless, DFS shows a modest runtime advantage (typically under 200 ms). 
This is attributed to two factors. 
First, in BFS, each tree level is processed using Rayon's parallel iterators. 
While this simplifies parallelization, it introduces scheduling overhead: threads are spawned 
at each level of the tree, causing latency from synchronization and load balancing. 
In contrast, DFS implementation statically partitions the binary tree into independent subtrees, each assigned to a thread before the recursion. 
This allows threads to proceed without interruption, thus improving efficiency.
Second, DFS achieves better cache locality compared to BFS. 
As discussed before, BFS traversal has less local reuse and accesses cross-level caches to store a large amount of data.
In contrast, DFS naturally promotes sequential and localized memory access due to its recursive structure, allowing intermediate data to be consumed quickly.
This increases the likelihood of cache hits
and leads to more efficient execution for large workloads.
While these advantages do not drastically change the overall asymptotic runtime, they could contribute to a consistent latency difference in practice.

Overall, in such compute-bound scenarios, traversal strategy has limited impact.
It matters more in the memory- or bandwidth-bound cases, particularly with hardware accelerators.

\subsection{MTU Parallel Execution and Speedup}

We now evaluate MTU performance across different numbers of processing elements (PEs), traversal strategies, and memory bandwidth constraints, ranging from DDR-level (64~GB/s) to HBM-scale (1024~GB/s). 
As in \autoref{sec: CPU perf}, we consider four representative ZKP workloads: Build MLE, MLE evaluation, Product MLE, and Merkle Tree commitment, each with a workload size of $2^{20}$. 
\autoref{fig:hardware_runtime} shows the runtime results under these varying configurations. 
For the BFS traversal, each PE operates independently on nodes of the same tree level. 
In contrast, in the DFS traversal, each PE processes a disjoint subtree (see \autoref{fig:build_mle} and \ref{fig:MLE_eval}), and a final PE merges the results. 
The MTU's Hybrid traversal combines BFS
and DFS 
to balance memory reuse and parallelism.
We start from two PEs as the MTU needs at least two PEs to support Hybrid traversal.

\textbf{Traversal Comparison.}  
We first analyze the difference between traversal methods. 
Under constrained bandwidth (e.g., 64 GB/s) and low parallelism (2 PEs), both DFS and Hybrid traversals significantly outperform BFS. 
This is because BFS reads and writes each tree level to off-chip memory, making it heavily bandwidth-bound even with a single PE. 
In contrast, DFS and Hybrid traversals retain intermediate values on-chip and only access off-chip memory for the initial input (or final output in the case of Build MLE), avoiding bandwidth-induced stalls.
An exception is 
the Product MLE workload, where
every tree level (including the intermediate levels) must output results.
Therefore, even in DFS and Hybrid modes, performance becomes bandwidth-limited due to the increased off-chip traffic for outputs storing.
On the other hand, in a high-bandwidth regime (1024 GB/s) with 2 PEs, traversal choice makes little difference, as all variants become compute-bound by the PE.
For workloads except Product MLE, DFS and Hybrid traversals achieve nearly 3$\times$ speedup over BFS, with sufficient bandwidth and compute resources.

\begin{figure}[t]
\centerline{\includegraphics[width=1.02\columnwidth]{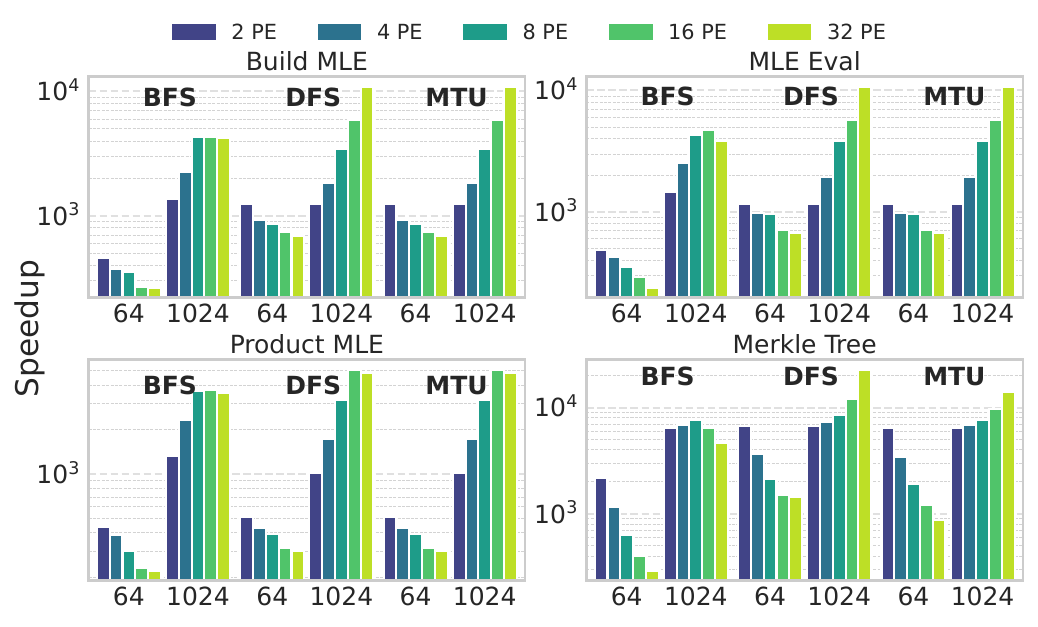}}
\caption{Hardware speedup over CPU baseline on workloads (each size $2^{20}$), grouped by different traversals. 
Each group shows DDR (low) and HBM-level (high) bandwidth (GB/s).
We start from 2 PEs to perform Hybrid traversal.}
\label{fig:hardware_speed_up}
\end{figure}

\textbf{Scalability with PE Parallelism.}  
The difference between traversals can be further evaluated by scaling the number of PEs under both bandwidth constraints.
At 64 GB/s, increasing the PE count yields little benefit across all traversal methods due to memory bandwidth saturation. 
With 1024 GB/s, the performance scales with PE count; however, BFS saturates quickly 
(plateauing after 16 PEs)
because of its bandwidth-heavy nature.
Product MLE exhibits similar limitations after 16 PEs, even under DFS or Hybrid traversal, since its per-level output demands high sustained bandwidth. 
For the other workloads, DFS and Hybrid traversals scale more effectively beyond 16 PEs, exhibiting better bandwidth efficiency.

Although DFS and Hybrid deliver comparable performance, a key limitation of DFS is its assumption that its input is divided into disjoint subtrees. 
DFS may be unsuitable when upstream modules deliver inputs in a contiguous, streaming fashion. 
In contrast, Hybrid traversal supports both parallelism and continuous input indexing, making it more practical in pipeline-integrated systems.

Our evaluation also reveals the significant impact of memory bandwidth on performance scalability. 
With 32 PEs, the system becomes memory-bound -- 
performance is limited
unless sufficient bandwidth is provisioned. 
As shown in \autoref{fig:hardware_runtime}, increasing the bandwidth from 64 GB/s to 1024 GB/s enables near-linear scaling, as the additional bandwidth alleviates the data movement bottleneck and allows the hardware to maintain high utilization across all PEs.

\textbf{Speedup over CPU.}
We compare MTU performance against CPU baselines under matched PE/thread configurations. 
For the CPU, we select the best-performing traversal method as the baseline. 
For MTU evaluations, we use two bandwidth levels: 64 GB/s, which is comparable to typical DDR4 memory bandwidth on CPUs, and 1024~GB/s, representing HBM-class bandwidth to illustrate the performance ceiling of MTU.
Speedup results are shown in \autoref{fig:hardware_speed_up}. 
The observed speedup trends mirror the hardware runtime patterns as discussed earlier. 
Notably, under the 64~GB/s bandwidth constraint, MTU's speedup diminishes as the number of PEs increases because the hardware is memory-bound.
Meanwhile, CPU performance continues to scale with increasing threads, as the workload remains compute-bound (see \autoref{fig:CPU_runtime}), so the relative hardware speedup decreases as the CPU improves with additional threads.

Overall, under CPU-class bandwidths (64~GB/s), MTU with Hybrid traversal achieves an average speedup of $1478\times$ over the CPU baseline. 
Under higher bandwidths (1024~GB/s), MTU scales much better, reaching up to $9440\times$ average speedup because it can 
exploit high degrees of parallelism and sufficient memory bandwidth.

\begin{figure}[t]
\centerline{\includegraphics[width=.79\columnwidth]{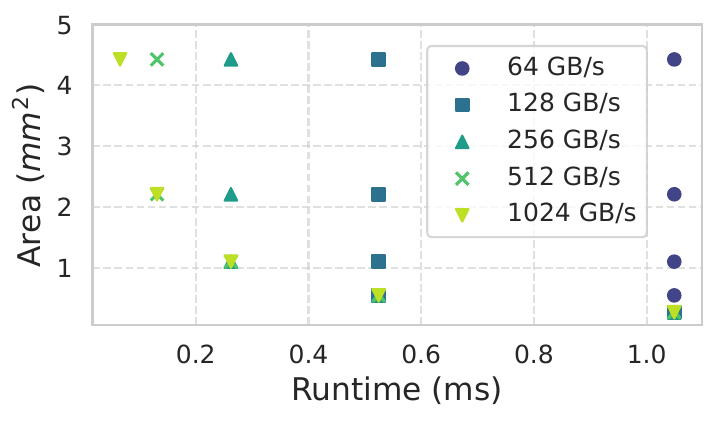}}
\caption{Runtime-area design space of MTU using Hybrid traversal on the Merkle Tree Commit workload. 
Each point represents a design configuration with a number of PEs and bandwidth. 
The Pareto frontier illustrates the trade-off between runtime and area across different bandwidth settings.}
\label{fig:Pareto}
\end{figure}

\subsection{Area, Power and Pareto Space Analysis}

\autoref{fig:Pareto} illustrates the runtime versus area design space of MTU under various bandwidth configurations and PE counts, using the Hybrid traversal scheme. 
We sweep a range of bandwidths (from 64~GB/s to 1024~GB/s) and evaluate the design points using the Merkle Tree commit workload with a problem size of $2^{20}$. 
Other workloads have similar Pareto trends as well.

The figure demonstrates a clear trend: increasing available bandwidth reduces runtime and shifts the Pareto front to the left. 
For large-area configurations (i.e., higher PE counts), additional bandwidth yields greater performance gains. 
However, for designs below 3~mm$^2$, the Pareto fronts across several bandwidth levels start to overlap, indicating that high bandwidth is not always necessary. 
For instance, a design slightly larger than 1~mm$^2$ achieves comparable performance under 256~GB/s and 1024~GB/s, suggesting that modest bandwidth suffices for smaller configurations.

\textbf{Prospect of MTU.}
\autoref{tab:area_power} summarizes the area and 
thermal design power of a representative MTU configuration with 32 PEs. 
MTU exhibits a compact footprint, and its area can be further reduced under limited bandwidth settings by scaling down the number of PEs. 
This 
enables MTU to be embedded within a larger SoC or included as part of a chiplet-based system while sharing a common HBM PHY interface for off-chip memory access.
For example, MTU is integrated into the zkSpeed accelerator~\cite{daftardar2025zkSpeed} to support ZKP generation. 
Beyond this case, MTU can be deployed to support potential SumCheck accelerators for MLE-related computation, or repurposed as a polynomial commitment engine within broader ZKP hardware stacks.
Its versatility and compactness make MTU a promising building block for modular ZKP accelerator architectures across a range of proof systems and deployment platforms.

\section{Related work}
\label{sec:Related}

Several prior works have explored hardware acceleration for ZKP protocols across various platforms and protocol families. 
\textbf{SZKP} \cite{szkp} accelerates the Groth16 protocol by optimizing MSM using Pippenger’s algorithm and dedicated scheduling strategies. 
MSM over elliptic curves serves as a PCS alternative to Merkle Tree-based commitments, offering smaller proof sizes but at higher computational cost. 
\textbf{NoCap}~\cite{nocap} proposes a vector-based processor targeting Spartan+Orion ZKP protocols. 
These protocols rely on 
SumCheck and Merkle Tree as core building blocks. 
While NoCap supports SumCheck, it does not provide specialization for SumCheck subroutines such as Build MLE or MLE evaluation. 
Merkle Tree computation is mapped to the vector engine directly using level-order BFS traversal.
\textbf{zkSpeed}~\cite{daftardar2025zkSpeed} is an ASIC that accelerates HyperPlonk proof generation end-to-end, with dedicated modules for MSM, SumCheck and its associated subroutines. 
\textbf{UniZK}~\cite{unizk} presents a systolic-array-based architecture that accelerates NTT and Merkle Tree. 
It optimized the Merkle Tree by partitioning a tree into subtrees, and executing each subtree on-chip with BFS.

\begin{table}[t]
\centering
\caption{Area/power breakdown of MTU hardware (32 PEs).
Misc includes the control and scheduler cost.
}
\label{tab:area_power}
\vspace{-0.8em}
\resizebox{0.72\columnwidth}{!}{
\setlength{\tabcolsep}{2mm}{
\begin{tabular}{lrr}
\hline
Component          & Area (mm$^2$)  & TDP (W)        \\ \hline
Modulus Ops        & 4.427          & 6.886          \\
SHA3               & 0.192          & 0.320          \\
{Misc}               & 0.011          & 0.010          \\
Memory             & 0.067          & 0.003          \\
\textbf{Total MTU} & \textbf{4.697} & \textbf{7.219} \\ \hline
HBM2 PHY           & 14.90          & 0.225          \\ \hline
\end{tabular}
}
}
\end{table}

Several prior works have explored traversal optimizations 
to address application-specific constraints. 
Optimizations in~\cite{knecht2014space,berman2007optimal} aim to improve efficiency when the total workload size is unknown and reduce redundant hash computations by speculating on future authentication paths. 
Other hardware-focused work~\cite{shadab2023hmt} improves Merkle Tree updates by reducing unnecessary memory traffic through cache-aware designs. 
For non-cryptographic workloads, for example,~\cite{anghel2019breadth} proposes switching between BFS and DFS during Random Forest training depending on the active sample size. 
While these optimizations are effective for their target applications, a unified hardware evaluation of traversal strategies is not provided for the diverse tree-based workloads in modern ZKP protocols.

\section{Conclusion}
\label{sec:conclusion}

{We present a systematic evaluation of binary tree-based ZKP workloads, analyzing traversal strategies and hardware trade-offs on the MTU architecture. Our results show that traversal strategy critically affects performance, with the proposed Hybrid Traversal offering significant gains in runtime and memory usage. MTU achieves up to 1478$\times$ speedup over CPU baselines under DDR-level bandwidth. These results emphasize MTU's practicality and portability as a general-purpose building block for modular ZKP hardware.}

\begin{acks}

This research was developed with funding in part from the NSF CAREER award \#2340137, the NSF CIRC Grand award \#2450539, and from DARPA under the Hybrid Electromagnetic side-channel and Interactive-proof Methods to Detect and Amend LogicaL Rifts (HEIMDALLR) program, grant number HR0011-25-C-0300.

\end{acks}

\bibliographystyle{ACM-Reference-Format}
\bibliography{sample-base}


\begin{thebibliography}{43}


\ifx \showCODEN    \undefined \def \showCODEN     #1{\unskip}     \fi
\ifx \showISBNx    \undefined \def \showISBNx     #1{\unskip}     \fi
\ifx \showISBNxiii \undefined \def \showISBNxiii  #1{\unskip}     \fi
\ifx \showISSN     \undefined \def \showISSN      #1{\unskip}     \fi
\ifx \showLCCN     \undefined \def \showLCCN      #1{\unskip}     \fi
\ifx \shownote     \undefined \def \shownote      #1{#1}          \fi
\ifx \showarticletitle \undefined \def \showarticletitle #1{#1}   \fi
\ifx \showURL      \undefined \def \showURL       {\relax}        \fi
\providecommand\bibfield[2]{#2}
\providecommand\bibinfo[2]{#2}
\providecommand\natexlab[1]{#1}
\providecommand\showeprint[2][]{arXiv:#2}

\bibitem[hbm(2023)]%
        {hbm3}
 \bibinfo{year}{2023}\natexlab{}.
\newblock \bibinfo{booktitle}{\emph{High Bandwidth Memory DRAM (HBM3)}}.
\newblock \bibinfo{type}{Technical Report} JESD238A. \bibinfo{institution}{JEDEC}.
\newblock
\newblock
\shownote{\url{https://www.jedec.org/standards-documents/docs/jesd238a}}.


\bibitem[Anghel et~al\mbox{.}(2019)]%
        {anghel2019breadth}
\bibfield{author}{\bibinfo{person}{Andreea~Simona Anghel}, \bibinfo{person}{Nikolas Ioannou}, \bibinfo{person}{Thomas Parnell}, \bibinfo{person}{Nikolaos Papandreou}, \bibinfo{person}{Celestine D{\"u}nner}, {and} \bibinfo{person}{Haris Pozidis}.} \bibinfo{year}{2019}\natexlab{}.
\newblock \showarticletitle{Breadth-first, Depth-next Training of Random Forests}. In \bibinfo{booktitle}{\emph{Annual Conference on Neural Information Processing Systems}}.
\newblock


\bibitem[arkworks(2022)]%
        {arkworks}
\bibfield{author}{\bibinfo{person}{arkworks}.} \bibinfo{year}{2022}\natexlab{}.
\newblock \bibinfo{booktitle}{\emph{\texttt{arkworks} zkSNARK ecosystem}}.
\newblock
\urldef\tempurl%
\url{https://arkworks.rs}
\showURL{%
\tempurl}


\bibitem[Ben-Sasson et~al\mbox{.}(2014)]%
        {zerocash}
\bibfield{author}{\bibinfo{person}{Eli Ben-Sasson}, \bibinfo{person}{Alessandro Chiesa}, \bibinfo{person}{Christina Garman}, \bibinfo{person}{Matthew Green}, \bibinfo{person}{Ian Miers}, \bibinfo{person}{Eran Tromer}, {and} \bibinfo{person}{Madars Virza}.} \bibinfo{year}{2014}\natexlab{}.
\newblock \showarticletitle{Zerocash: Decentralized Anonymous Payments from Bitcoin}.
\newblock  (\bibinfo{year}{2014}).
\newblock


\bibitem[Berman et~al\mbox{.}(2007)]%
        {berman2007optimal}
\bibfield{author}{\bibinfo{person}{Piotr Berman}, \bibinfo{person}{Marek Karpinski}, {and} \bibinfo{person}{Yakov Nekrich}.} \bibinfo{year}{2007}\natexlab{}.
\newblock \showarticletitle{Optimal trade-off for Merkle tree traversal}.
\newblock \bibinfo{journal}{\emph{Theoretical Computer Science}} \bibinfo{volume}{372}, \bibinfo{number}{1} (\bibinfo{year}{2007}), \bibinfo{pages}{26--36}.
\newblock


\bibitem[Bertoni et~al\mbox{.}(2009)]%
        {SHA3keccak}
\bibfield{author}{\bibinfo{person}{Guido Bertoni}, \bibinfo{person}{Joan Daemen}, \bibinfo{person}{Micha{\"e}l Peeters}, {and} \bibinfo{person}{Gilles Van~Assche}.} \bibinfo{year}{2009}\natexlab{}.
\newblock \showarticletitle{Keccak specifications}.
\newblock \bibinfo{journal}{\emph{Submission to nist (round 2)}} \bibinfo{volume}{3}, \bibinfo{number}{30} (\bibinfo{year}{2009}), \bibinfo{pages}{320--337}.
\newblock


\bibitem[Blumberg et~al\mbox{.}(2014)]%
        {blumberg2014verifiable}
\bibfield{author}{\bibinfo{person}{Andrew~J Blumberg}, \bibinfo{person}{Justin Thaler}, \bibinfo{person}{Victor Vu}, {and} \bibinfo{person}{Michael Walfish}.} \bibinfo{year}{2014}\natexlab{}.
\newblock \showarticletitle{Verifiable computation using multiple provers}.
\newblock \bibinfo{journal}{\emph{Cryptology ePrint Archive}} (\bibinfo{year}{2014}).
\newblock


\bibitem[Bonneau et~al\mbox{.}(2020)]%
        {bonneau2020mina}
\bibfield{author}{\bibinfo{person}{Joseph Bonneau}, \bibinfo{person}{Izaak Meckler}, \bibinfo{person}{Vanishree Rao}, {and} \bibinfo{person}{Evan Shapiro}.} \bibinfo{year}{2020}\natexlab{}.
\newblock \showarticletitle{Mina: Decentralized Cryptocurrency at Scale}.
\newblock  (\bibinfo{year}{2020}).
\newblock


\bibitem[by~Micron Technology~Inc.(2023)]%
        {DDR4_crucial}
\bibfield{author}{\bibinfo{person}{Crucial by Micron Technology~Inc.}} \bibinfo{year}{2023}\natexlab{}.
\newblock \bibinfo{title}{Ram Memory Speeds and Compatibility}.
\newblock
\urldef\tempurl%
\url{https://www.crucial.com/support/memory-speeds-compatability}
\showURL{%
\tempurl}


\bibitem[Campanelli et~al\mbox{.}(2019)]%
        {campanelli2019legosnark}
\bibfield{author}{\bibinfo{person}{Matteo Campanelli}, \bibinfo{person}{Dario Fiore}, {and} \bibinfo{person}{Ana{\"\i}s Querol}.} \bibinfo{year}{2019}\natexlab{}.
\newblock \showarticletitle{LegoSNARK: Modular Design and Composition of Succinct Zero-Knowledge Proofs}.
\newblock  (\bibinfo{year}{2019}).
\newblock


\bibitem[Chen et~al\mbox{.}(2022)]%
        {hyperplonk}
\bibfield{author}{\bibinfo{person}{Binyi Chen}, \bibinfo{person}{Benedikt Bünz}, \bibinfo{person}{Dan Boneh}, {and} \bibinfo{person}{Zhenfei Zhang}.} \bibinfo{year}{2022}\natexlab{}.
\newblock \bibinfo{title}{{HyperPlonk}: Plonk with Linear-Time Prover and High-Degree Custom Gates}.
\newblock \bibinfo{howpublished}{Cryptology {ePrint} Archive, Paper 2022/1355}.
\newblock
\urldef\tempurl%
\url{https://eprint.iacr.org/2022/1355}
\showURL{%
\tempurl}


\bibitem[Company(2021)]%
        {22nm_TSMC_web}
\bibfield{author}{\bibinfo{person}{Taiwan Semiconductor~Manufacturing Company}.} \bibinfo{year}{2021}\natexlab{}.
\newblock \bibinfo{title}{TSMC 22nm Technology}.
\newblock
\urldef\tempurl%
\url{https://www.tsmc.com/english/dedicatedFoundry/technology/logic/l_22nm}
\showURL{%
\tempurl}


\bibitem[Daftardar et~al\mbox{.}(2025a)]%
        {daftardar2025zkphire}
\bibfield{author}{\bibinfo{person}{Alhad Daftardar}, \bibinfo{person}{Jianqiao Mo}, \bibinfo{person}{Joey Ah-kiow}, \bibinfo{person}{Benedikt B{\"u}nz}, \bibinfo{person}{Siddharth Garg}, {and} \bibinfo{person}{Brandon Reagen}.} \bibinfo{year}{2025}\natexlab{a}.
\newblock \showarticletitle{zkPHIRE: A Programmable Accelerator for ZKPs over HIgh-degRee, Expressive Gates}.
\newblock \bibinfo{journal}{\emph{arXiv preprint arXiv:2508.16738}} (\bibinfo{year}{2025}).
\newblock


\bibitem[Daftardar et~al\mbox{.}(2025b)]%
        {daftardar2025zkSpeed}
\bibfield{author}{\bibinfo{person}{Alhad Daftardar}, \bibinfo{person}{Jianqiao Mo}, \bibinfo{person}{Joey Ah-kiow}, \bibinfo{person}{Benedikt B{\"u}nz}, \bibinfo{person}{Ramesh Karri}, \bibinfo{person}{Siddharth Garg}, {and} \bibinfo{person}{Brandon Reagen}.} \bibinfo{year}{2025}\natexlab{b}.
\newblock \showarticletitle{Need for zkSpeed: Accelerating HyperPlonk for Zero-Knowledge Proofs}. In \bibinfo{booktitle}{\emph{Proceedings of the 52nd Annual International Symposium on Computer Architecture}}. \bibinfo{pages}{1986--2001}.
\newblock


\bibitem[Daftardar et~al\mbox{.}(2024)]%
        {szkp}
\bibfield{author}{\bibinfo{person}{Alhad Daftardar}, \bibinfo{person}{Brandon Reagen}, {and} \bibinfo{person}{Siddharth Garg}.} \bibinfo{year}{2024}\natexlab{}.
\newblock \showarticletitle{Szkp: A scalable accelerator architecture for zero-knowledge proofs}. In \bibinfo{booktitle}{\emph{Proceedings of the 2024 International Conference on Parallel Architectures and Compilation Techniques}}. \bibinfo{pages}{271--283}.
\newblock


\bibitem[Erickson(2023)]%
        {erickson2023algorithms}
\bibfield{author}{\bibinfo{person}{Jeff Erickson}.} \bibinfo{year}{2023}\natexlab{}.
\newblock \bibinfo{booktitle}{\emph{Algorithms}}.
\newblock


\bibitem[Feng et~al\mbox{.}(2023)]%
        {10.1145/3617232.3624852}
\bibfield{author}{\bibinfo{person}{Boyuan Feng}, \bibinfo{person}{Zheng Wang}, \bibinfo{person}{Yuke Wang}, \bibinfo{person}{Shu Yang}, {and} \bibinfo{person}{Yufei Ding}.} \bibinfo{year}{2023}\natexlab{}.
\newblock \showarticletitle{ZENO: A Type-based Optimization Framework for Zero Knowledge Neural Network Inference}.
\newblock  (\bibinfo{year}{2023}).
\newblock


\bibitem[Gabizon et~al\mbox{.}(2019)]%
        {plonk}
\bibfield{author}{\bibinfo{person}{Ariel Gabizon}, \bibinfo{person}{Zachary~J. Williamson}, {and} \bibinfo{person}{Oana Ciobotaru}.} \bibinfo{year}{2019}\natexlab{}.
\newblock \bibinfo{title}{{PLONK}: Permutations over Lagrange-bases for Oecumenical Noninteractive arguments of Knowledge}.
\newblock \bibinfo{howpublished}{Cryptology {ePrint} Archive, Paper 2019/953}.
\newblock
\urldef\tempurl%
\url{https://eprint.iacr.org/2019/953}
\showURL{%
\tempurl}


\bibitem[Geng et~al\mbox{.}(2023)]%
        {geng2023privacy}
\bibfield{author}{\bibinfo{person}{Haoran Geng}, \bibinfo{person}{Jianqiao Mo}, \bibinfo{person}{Dayane Reis}, \bibinfo{person}{Jonathan Takeshita}, \bibinfo{person}{Taeho Jung}, \bibinfo{person}{Brandon Reagen}, \bibinfo{person}{Michael Niemier}, {and} \bibinfo{person}{Xiaobo~Sharon Hu}.} \bibinfo{year}{2023}\natexlab{}.
\newblock \showarticletitle{Privacy preserving in-memory computing engine}.
\newblock \bibinfo{journal}{\emph{arXiv preprint arXiv:2308.02648}} (\bibinfo{year}{2023}).
\newblock


\bibitem[Grassi et~al\mbox{.}(2019)]%
        {poseidon}
\bibfield{author}{\bibinfo{person}{Lorenzo Grassi}, \bibinfo{person}{Dmitry Khovratovich}, \bibinfo{person}{Christian Rechberger}, \bibinfo{person}{Arnab Roy}, {and} \bibinfo{person}{Markus Schofnegger}.} \bibinfo{year}{2019}\natexlab{}.
\newblock \showarticletitle{POSEIDON: A New Hash Function for Zero-Knowledge Proof Systems}.
\newblock  (\bibinfo{year}{2019}).
\newblock


\bibitem[Gully et~al\mbox{.}(2013)]%
        {intel_sha3}
\bibfield{author}{\bibinfo{person}{Sean Gully}, \bibinfo{person}{Vinodh Gopal}, \bibinfo{person}{Kirk Yap}, \bibinfo{person}{Wajdi Feghali}, \bibinfo{person}{Jim Guilford}, {and} \bibinfo{person}{Gil Wolrich}.} \bibinfo{year}{2013}\natexlab{}.
\newblock \showarticletitle{New Instructions Supporting the Secure Hash Algorithm on Intel Architecture Processors}.
\newblock \bibinfo{journal}{\emph{Tech. Rep.}} (\bibinfo{year}{2013}).
\newblock


\bibitem[Inc.(2020)]%
        {hbm2}
\bibfield{author}{\bibinfo{person}{Rambus Inc.}} \bibinfo{year}{2020}\natexlab{}.
\newblock \bibinfo{title}{White Paper: HBM2E and GDDR6: Memory Solutions for AI}.
\newblock


\bibitem[Intel(2025)]%
        {Intel_CPU}
\bibfield{author}{\bibinfo{person}{Intel}.} \bibinfo{year}{2025}\natexlab{}.
\newblock \bibinfo{title}{Intel xeon gold 5218 processor (22m cache, 2.30 ghz) 2025}.
\newblock
\urldef\tempurl%
\url{https://www.intel.com/content/www/us/en/products/sku/192444/intel-xeon-gold-5218-processor-22m-cache-2-30-ghz/specifications.html}
\showURL{%
\tempurl}


\bibitem[Kim et~al\mbox{.}(2021)]%
        {bts}
\bibfield{author}{\bibinfo{person}{Sangpyo Kim}, \bibinfo{person}{Jongmin Kim}, \bibinfo{person}{Michael~Jaemin Kim}, \bibinfo{person}{Wonkyung Jung}, \bibinfo{person}{Minsoo Rhu}, \bibinfo{person}{John Kim}, {and} \bibinfo{person}{Jung~Ho Ahn}.} \bibinfo{year}{2021}\natexlab{}.
\newblock \showarticletitle{BTS: An Accelerator for Bootstrappable Fully Homomorphic Encryption}.
\newblock \bibinfo{journal}{\emph{arXiv preprint arXiv:2112.15479}} (\bibinfo{year}{2021}).
\newblock


\bibitem[Knecht and Nicola(2013)]%
        {knecht2014space}
\bibfield{author}{\bibinfo{person}{Markus Knecht} {and} \bibinfo{person}{Carlo~U. Nicola}.} \bibinfo{year}{2013}\natexlab{}.
\newblock \showarticletitle{A space- and time-efficient Implementation of the Merkle Tree Traversal Algorithm}.
\newblock \bibinfo{journal}{\emph{IMVS Fokus Report}}  \bibinfo{volume}{7} (\bibinfo{year}{2013}), \bibinfo{pages}{35--39}.
\newblock
\showISSN{2296-4169}
\href{https://doi.org/11654/17789}{doi:\nolinkurl{11654/17789}}


\bibitem[Mo et~al\mbox{.}(2023a)]%
        {mo2023towards}
\bibfield{author}{\bibinfo{person}{Jianqiao Mo}, \bibinfo{person}{Karthik Garimella}, \bibinfo{person}{Negar Neda}, \bibinfo{person}{Austin Ebel}, {and} \bibinfo{person}{Brandon Reagen}.} \bibinfo{year}{2023}\natexlab{a}.
\newblock \showarticletitle{Towards fast and scalable private inference}. In \bibinfo{booktitle}{\emph{Proceedings of the 20th ACM International Conference on Computing Frontiers}}. \bibinfo{pages}{322--328}.
\newblock


\bibitem[Mo et~al\mbox{.}(2023b)]%
        {haac}
\bibfield{author}{\bibinfo{person}{Jianqiao Mo}, \bibinfo{person}{Jayanth Gopinath}, {and} \bibinfo{person}{Brandon Reagen}.} \bibinfo{year}{2023}\natexlab{b}.
\newblock \showarticletitle{Haac: A hardware-software co-design to accelerate garbled circuits}. In \bibinfo{booktitle}{\emph{Proceedings of the 50th Annual International Symposium on Computer Architecture}}. \bibinfo{pages}{1--13}.
\newblock


\bibitem[Mo and Reagen(2023)]%
        {moaccelerating}
\bibfield{author}{\bibinfo{person}{Jianqiao Mo} {and} \bibinfo{person}{Brandon Reagen}.} \bibinfo{year}{2023}\natexlab{}.
\newblock \showarticletitle{Accelerating Garbled Circuits by Hardware-Software Co-Design}.
\newblock \bibinfo{journal}{\emph{DISCC 2023 2nd Workshop on Data Integrity and Secure Cloud Computing}} (\bibinfo{year}{2023}).
\newblock


\bibitem[Nukada(2021)]%
        {DDR4}
\bibfield{author}{\bibinfo{person}{Akira Nukada}.} \bibinfo{year}{2021}\natexlab{}.
\newblock \showarticletitle{Performance Optimization of Allreduce Operation for Multi-GPU Systems}. In \bibinfo{booktitle}{\emph{2021 IEEE International Conference on Big Data (Big Data)}}. IEEE, \bibinfo{pages}{3107--3112}.
\newblock


\bibitem[{OpenCores}(2013)]%
        {opencores_sha3}
\bibfield{author}{\bibinfo{person}{{OpenCores}}.} \bibinfo{year}{2013}\natexlab{}.
\newblock \bibinfo{title}{SHA-3 IP Core}.
\newblock \bibinfo{howpublished}{\url{https://opencores.org/projects/sha3}}.
\newblock


\bibitem[Samardzic et~al\mbox{.}(2024)]%
        {nocap}
\bibfield{author}{\bibinfo{person}{Nikola Samardzic}, \bibinfo{person}{Simon Langowski}, \bibinfo{person}{Srinivas Devadas}, {and} \bibinfo{person}{Daniel Sanchez}.} \bibinfo{year}{2024}\natexlab{}.
\newblock \bibinfo{title}{Accelerating Zero-Knowledge Proofs Through Hardware-Algorithm Co-Design}.
\newblock \bibinfo{howpublished}{Preprint}.
\newblock
\newblock
\shownote{https://people.csail.mit.edu/devadas/pubs/micro24\_nocap.pdf}.


\bibitem[Schlachter and Drake(2019)]%
        {micron_ddr5}
\bibfield{author}{\bibinfo{person}{Scott Schlachter} {and} \bibinfo{person}{Brian Drake}.} \bibinfo{year}{2019}\natexlab{}.
\newblock \showarticletitle{Introducing Micron{\textregistered} DDR5 SDRAM: More than a generational update}.
\newblock \bibinfo{journal}{\emph{XP055844818}}  \bibinfo{volume}{31} (\bibinfo{year}{2019}), \bibinfo{pages}{6}.
\newblock


\bibitem[Setty(2020)]%
        {spartan}
\bibfield{author}{\bibinfo{person}{Srinath Setty}.} \bibinfo{year}{2020}\natexlab{}.
\newblock \showarticletitle{Spartan: Efficient and general-purpose zkSNARKs without trusted setup}. In \bibinfo{booktitle}{\emph{Annual International Cryptology Conference}}. Springer, \bibinfo{pages}{704--737}.
\newblock


\bibitem[Setty and Lee(2020)]%
        {setty2020quarks}
\bibfield{author}{\bibinfo{person}{Srinath Setty} {and} \bibinfo{person}{Jonathan Lee}.} \bibinfo{year}{2020}\natexlab{}.
\newblock \showarticletitle{Quarks: Quadruple-efficient transparent zkSNARKs}.
\newblock \bibinfo{journal}{\emph{Cryptology ePrint Archive}} (\bibinfo{year}{2020}).
\newblock


\bibitem[Shadab et~al\mbox{.}(2022)]%
        {shadab2023hmt}
\bibfield{author}{\bibinfo{person}{Rakin~Muhammad Shadab}, \bibinfo{person}{Yu Zou}, \bibinfo{person}{Sanjay Gandham}, \bibinfo{person}{Amro Awad}, {and} \bibinfo{person}{Mingjie Lin}.} \bibinfo{year}{2022}\natexlab{}.
\newblock \showarticletitle{HMT: A Hardware-Centric Hybrid Bonsai Merkle Tree Algorithm for High-Performance Authentication}.
\newblock \bibinfo{journal}{\emph{arXiv preprint arXiv:2204.08976}} (\bibinfo{year}{2022}).
\newblock


\bibitem[Thaler(2022)]%
        {thaler_proofs_args_zk}
\bibfield{author}{\bibinfo{person}{Justin Thaler}.} \bibinfo{year}{2022}\natexlab{}.
\newblock \bibinfo{booktitle}{\emph{Proofs, Arguments, and Zero-Knowledge}}.
\newblock
\urldef\tempurl%
\url{https://people.cs.georgetown.edu/jthaler/ProofsArgsAndZK.pdf}
\showURL{%
\tempurl}


\bibitem[Wahby et~al\mbox{.}(2017)]%
        {cryptoeprint:2017/1132}
\bibfield{author}{\bibinfo{person}{Riad~S. Wahby}, \bibinfo{person}{Ioanna Tzialla}, \bibinfo{person}{abhi shelat}, \bibinfo{person}{Justin Thaler}, {and} \bibinfo{person}{Michael Walfish}.} \bibinfo{year}{2017}\natexlab{}.
\newblock \bibinfo{title}{Doubly-efficient {zkSNARKs} without trusted setup}.
\newblock \bibinfo{howpublished}{Cryptology {ePrint} Archive, Paper 2017/1132}.
\newblock
\urldef\tempurl%
\url{https://eprint.iacr.org/2017/1132}
\showURL{%
\tempurl}


\bibitem[Wang and Gao(2025)]%
        {unizk}
\bibfield{author}{\bibinfo{person}{Cheng Wang} {and} \bibinfo{person}{Mingyu Gao}.} \bibinfo{year}{2025}\natexlab{}.
\newblock \showarticletitle{UniZK: Accelerating Zero-Knowledge Proof with Unified Hardware and Flexible Kernel Mapping}.
\newblock  (\bibinfo{year}{2025}).
\newblock


\bibitem[Wu et~al\mbox{.}(2016)]%
        {7nm_TSMC_paper}
\bibfield{author}{\bibinfo{person}{Shien-Yang Wu}, \bibinfo{person}{CY Lin}, \bibinfo{person}{MC Chiang}, \bibinfo{person}{JJ Liaw}, \bibinfo{person}{JY Cheng}, \bibinfo{person}{SH Yang}, \bibinfo{person}{CH Tsai}, \bibinfo{person}{PN Chen}, \bibinfo{person}{T Miyashita}, \bibinfo{person}{CH Chang}, {et~al\mbox{.}}} \bibinfo{year}{2016}\natexlab{}.
\newblock \showarticletitle{A 7nm CMOS platform technology featuring 4 th generation FinFET transistors with a 0.027 um 2 high density 6-T SRAM cell for mobile SoC applications}. In \bibinfo{booktitle}{\emph{2016 IEEE IEDM}}. IEEE, \bibinfo{pages}{2--6}.
\newblock


\bibitem[Wu et~al\mbox{.}(2013)]%
        {16nm_TSMC_paper}
\bibfield{author}{\bibinfo{person}{Shien-Yang Wu}, \bibinfo{person}{Colin~Yu Lin}, \bibinfo{person}{MC Chiang}, \bibinfo{person}{JJ Liaw}, \bibinfo{person}{JY Cheng}, \bibinfo{person}{SH Yang}, \bibinfo{person}{Ming Liang}, \bibinfo{person}{Tadakazu Miyashita}, \bibinfo{person}{CH Tsai}, \bibinfo{person}{BC Hsu}, {et~al\mbox{.}}} \bibinfo{year}{2013}\natexlab{}.
\newblock \showarticletitle{A 16nm FinFET CMOS technology for mobile SoC and computing applications}. In \bibinfo{booktitle}{\emph{2013 IEEE International Electron Devices Meeting}}. IEEE, \bibinfo{pages}{9--1}.
\newblock


\bibitem[Xie et~al\mbox{.}(2019)]%
        {cryptoeprint:2019/317}
\bibfield{author}{\bibinfo{person}{Tiancheng Xie}, \bibinfo{person}{Jiaheng Zhang}, \bibinfo{person}{Yupeng Zhang}, \bibinfo{person}{Charalampos Papamanthou}, {and} \bibinfo{person}{Dawn Song}.} \bibinfo{year}{2019}\natexlab{}.
\newblock \showarticletitle{Libra: Succinct Zero-Knowledge Proofs with Optimal Prover Computation}.
\newblock  (\bibinfo{year}{2019}).
\newblock


\bibitem[Xie et~al\mbox{.}(2022)]%
        {orion}
\bibfield{author}{\bibinfo{person}{Tiancheng Xie}, \bibinfo{person}{Yupeng Zhang}, {and} \bibinfo{person}{Dawn Song}.} \bibinfo{year}{2022}\natexlab{}.
\newblock \showarticletitle{Orion: Zero Knowledge Proof with Linear Prover Time}.
\newblock \bibinfo{journal}{\emph{Cryptology ePrint Archive}} (\bibinfo{year}{2022}).
\newblock


\bibitem[Zhang et~al\mbox{.}(2022)]%
        {hyperplonkespresso}
\bibfield{author}{\bibinfo{person}{Zhenfei Zhang}, \bibinfo{person}{Binyi Chen}, \bibinfo{person}{Benedikt B\"unz}, {and} \bibinfo{person}{Alex Xiong}.} \bibinfo{year}{2022}\natexlab{}.
\newblock \bibinfo{title}{Hyperplonk Implementation}.
\newblock
\urldef\tempurl%
\url{https://github.com/EspressoSystems/hyperplonk}
\showURL{%
\tempurl}


\end{thebibliography}

\end{document}